\documentclass[superscriptaddress,longbibliography,amsmath,amssymb,aps,prx,notitlepage,twocolumn,floatfix,nofootinbib]{revtex4-2}
\usepackage{graphicx,xcolor}
\usepackage{amsmath,amssymb}
\usepackage{braket}
\usepackage{lipsum}
\usepackage[colorlinks=true, linkcolor=blue, citecolor=blue, urlcolor=blue]{hyperref}
\let\oldref\ref
\renewcommand{\ref}[1]{\textcolor{blue}{[\oldref{#1}]}}
\usepackage{cleveref}

\renewcommand{\vec}[1]{\boldsymbol{#1}}

\usepackage{dcolumn}
\usepackage{bbold}

\usepackage{color}
\definecolor{darkgreen}{rgb}{0.10, 0.65, 0.10}

\newcommand{\LCSO}{Li$_3$Co$_2$SbO$_6$}
\newcommand{\NCSO}{Na$_3$Co$_2$SbO$_6$}
\newcommand{\NCTO}{Na$_2$Co$_2$TeO$_6$}

\AtBeginDocument{%
    \newwrite\bibnotes
    \def\bibnotesext{Notes.bib}
    \immediate\openout\bibnotes=\jobname\bibnotesext
    \immediate\write\bibnotes{@CONTROL{REVTEX41Control}}
    \immediate\write\bibnotes{@CONTROL{%
    apsrev41Control,author="08",editor="1",pages="1",title="0",year="1"}}
     \if@filesw
     \immediate\write\@auxout{\string\citation{apsrev41Control}}%
    \fi
}%

\begin{document}

\title{Inelastic neutron scattering of the layered Kitaev ferromagnet \LCSO}

\author{Abdul Basit}
\affiliation{School of Physics, University of Melbourne, Parkville, VIC 3010, Australia}

\author{Richard A.\ Mole}
\affiliation{Australian Nuclear Science and Technology Organisation (ANSTO),
New Illawarra Road, Lucas Heights, NSW 2234, Australia}

\author{Alex J. Brown}
\affiliation{School of Chemistry, The University of Sydney, Sydney, NSW 2006, Australia}
\affiliation{School of Engineering, University of Warwick, Coventry, CV4 7AL, United Kingdom}

\author{Jiatu Liu}
\affiliation{School of Chemistry, The University of Sydney, Sydney, NSW 2006, Australia}
\affiliation{PETRA III, DESY, Geb.\,48f, Notkestraße 85, 22607 Hamburg, Germany}

\author{Chris D.\ Ling}
\affiliation{School of Chemistry, The University of Sydney, Sydney, NSW 2006, Australia}

\author{Stephan Rachel} 
\affiliation{School of Physics, University of Melbourne, Parkville, VIC 3010, Australia}

\date{\today}


\newcommand{\ncto}{Na$_2$Co$_2$TeO$_6$}
\newcommand{\ncso}{Na$_3$Co$_2$SbO$_6$}
\newcommand{\lcso}{Li$_3$Co$_2$SbO$_6$}


\begin{abstract}
Cobalt-based quantum magnets forming layered honeycomb arrangements have attracted much attention recently, as they are considered as a potential platform for materials with exotic Kitaev spin exchange. Amongst the discussed candidate materials are Na$_3$Co$_2$SbO$_6$ and Na$_2$Co$_2$TeO$_6$, both possessing a low-temperature ground state with some form of antiferromagnetic ordering (zigzag or triple-$Q$), similar to Na$_2$IrO$_3$ and $\alpha$-RuCl$_3$. Here we report inelastic neutron scattering experiments on the quantum magnet Li$_3$Co$_2$SbO$_6$, which features ferromagnetic honeycomb planes with opposite magnetizations in neighboring planes. By comparing with linear spin wave theory, we show that the magnetic properties of Li$_3$Co$_2$SbO$_6$ can be well-modelled by an extended Kitaev--Heisenberg model, establishing it as a Kitaev-ferromagnet, or more specifically, as a Kitaev A-type antiferromagnet. Our analysis is complemented by magnetic field measurements and simulations. 
\end{abstract}
\maketitle

\section{Introduction}
\label{sec:intro}
Over the past few decades, the search for quantum spin liquids\,\cite{Balents_2010, Lucile_2016} has attracted substantial attention within the condensed matter physics community. A spin liquid is a highly entangled phase of a magnetic system that does not possess any conventional long-range magnetic order. In contrast to most magnetic systems, which develop ordered ground states upon cooling, a quantum spin liquid evades such ordering entirely due to the absence of any symmetry breaking and retains its fluctuating, highly entangled character even at zero temperature. The notion of a spin liquid was originally introduced by Anderson within the resonating valence bond framework\,\cite{anderson73mrb153}. In this context, he proposed a solution to an antiferromagnetic spin model on the triangular lattice with nearest-neighbor interactions. The proposed ground state forms a highly degenerate manifold comprising all possible configurations of spin-singlet pairings among lattice sites, capturing the intrinsically resonant behavior of the resonating valence bond state. The disordered nature of the ground state along with the (non-local) formation of spin-singlet pairs imply that despite having no long-range magnetic order, the system exhibits entanglement on a macroscopic scale. 

Although Anderson’s proposal was innovative, subsequent studies established that the ground state of the antiferromagnetic Heisenberg model on the triangular lattice exhibits long-range magnetic order rather than a spin liquid phase. In contrast, a paradigmatic model supporting a true quantum spin liquid ground state was introduced many years later by Kitaev. In his seminal work\,\cite{Anyons_in_Kitaev}, Kitaev solved a bond-dependent, Ising-type spin model on the honeycomb lattice through a Majorana fermion decomposition. The solution provides a natural setting for fractionalization, in which the spin degrees of freedom separate into emergent quasiparticles consisting of Majorana fermions and static $\mathbb{Z}_2$ gauge fluxes (visons). The Kitaev model is the first spin Hamiltonian which possesses an exactly solvable quantum spin liquid ground state.
Kitaev's work and the hope of finding such a quantum spin liquid experimentally initiated the search and discovery of a whole class of materials that exhibits this bond-dependent spin-exchange. Kitaev candidate materials are usually identified as Mott insulators (or spin-orbit assisted Mott insulators) as Kitaev spin exchange leads to a high degree of frustration\,\cite{Jackelli_Khaliulin_khomskii, William_witczak_2014, Hwan_2015, GRau_2016, Motome_kitaev_design,liu_towards_2021, Kim_2022, Kitaev_materials_review_art}. 

To date, all Kitaev candidate materials have turned out to be magnetically ordered at low temperatures, revealing the presence of other conventional spin interactions which prevent the system from forming a true spin liquid ground state. There is hope, however, that some of the materials might be in close proximity to the sought-after Kitaev spin liquid phase\,\cite{Alpha-RuCl3_2, H_liu_towards_metal}. The effective spin interactions are expected to depend sensitively on external parameters such as magnetic field or applied pressure \cite{Alpha_RuCl3_1, liu_towards_2021, Pressure_tuning}, raising the possibility of tuning these systems into a Kitaev spin liquid ground state. For most materials it remains somewhat puzzling how large the various spin interactions are. Experimental verification is difficult and typically requires fitting model-derived observables to experimental data. Examples of experimental methods include spin susceptibility measurements\,\cite{lampen-kelley-18prb100403, dasMagneticAnisotropyAlkali2019, PhysRevB.102.224411}, Raman spectroscopy \cite{Raman_kitaev_1, Raman_kitaev_2, guptaRamanSignaturesStrong2016} and inelastic neutron scattering (INS)\,\cite{Alpha-RuCl3_2, INS_RuCl3_2, INS_RuCl3_3, Na2IrO3_4, songvilay-20prb224429, kim_antiferromagnetic_2021,  Alaric_etal_new,  chen_topological_2018, chen_magnetic_2020, CrCl3_1, NCTO_new_1, NCTO_new_2}.
In addition to these, theoretical techniques such as {\it ab initio} \cite{winter_models_2017, ab-initio_1, ab-initio_2,  ab_initio_liu,DFT_LSCO_imag} and {\it quantum-chemistry} methods \cite{ab-initio_3, NCSO_pritam, LCSO_quantum_chemistry_LH} can offer valuable insight about the underlying exchange interactions.

Jackeli and Khalliulin suggested that Kitaev spin exchange could be realized in the honeycomb iridate $\text{Na}_2\text{IrO}_3$\,\cite{Jackelli_Khaliulin_khomskii}. Subsequent experiments found the material to be magnetically ordered at low temperatures ($T_c \approx15 $K), but also to be compatible with non-negligible Kitaev spin exchange\,\cite{Na2IrO3_1, Na2IrO3_2, Na2IrO3_3, Na2IrO3_4}. Further attention was drawn to the sister compound $\alpha$-$\text{Li}_2\text{IrO}_3$ by both theoretical\,\cite{Li2IrO3_1,reuther-14prb100405, Li2IrO3_2} and experimental work\,\cite{li2irO3_exp1,li2irO3_exp2}.
%
%
Another compound that has sparked and significantly influenced the research activity is
$\alpha$-$\text{Ru}\text{Cl}_3$. Structurally, it consists of layers of edge-sharing $\text{RuCl}_6$ octahedra forming a nearly ideal honeycomb lattice. The $\text{Ru}^{3+}$ ions possess effective spin-1/2 moments due to strong spin-orbit coupling, a key ingredient for Kitaev interactions. Although $\alpha$-RuCl$_3$ develops long-range antiferromagnetic order at low temperatures ($T_c \approx 7$ K), experimental evidence of fractionalized excitations consistent with the Kitaev model has been reported\,\cite{Alpha-RuCl3_2, RuCl3_3, Alpha_RuCl3_4, lampen-kelley-18prb100403, Alpha_RuCl3_1}. When subjected to a sufficiently strong external magnetic field, the magnetic order is suppressed, giving rise to a quantum-disordered regime that exhibits characteristics reminiscent of a quantum spin liquid\,\cite{Therm_cond_RuCl3_2, Thermal_cond_RuCl3_1_new}.

In recent years, honeycomb cobaltates were proposed to expand the pool of candidate materials\,\cite{liu_towards_2021, sano_etal_2018, H_liu_towards_metal, Kim_2022}. They represent a group of layered transition metal oxides, where cobalt ions effectively form a honeycomb lattice\,\cite{liu_pseudospin_2018,liu_towards_2021}. Non-negligible spin-orbit coupling mechanisms, along with some structural effects originating from  trigonal crystal fields, have been investigated in these compounds \cite{liu_towards_2021}. Investigated materials within this category include $\text{BaCo}_2(\text{AsO}_4)_2$ and $\text{BaCo}_2(\text{PO}_4)_2$\,\cite{BaCo_1, BaCo_2, BaCo_3} as well as \NCSO\,and \NCTO\,\cite{NCTO_new_4, Alaric_etal_new, songvilay-20prb224429, kim_antiferromagnetic_2021, NCTO_new_1, NCTO_new_2,NCTO_new_3, vavilova-23prb054411,dufault-23prb064405, arneth-24prbL140402,miao-24prb134431}. Previous works on $\text{Na}_{3}\text{Co}_2 \text{SbO}_6$ and $\text{Na}_{2}\text{Co}_2 \text{TeO}_6$\,\cite{Alaric_etal_new, songvilay-20prb224429, kim_antiferromagnetic_2021, NCTO_new_2} indicate the presence of strong Kitaev signatures within these materials despite their ground state spin configuration being magnetically ordered at low temperatures. Experiments on powder samples established them as zig-zag antiferromagnets similar to what was observed for $\text{Na}_{2}\text{IrO}_3$ and $\alpha$-$\text{Ru}\text{Cl}_3$\,\cite{Alaric_etal_new, songvilay-20prb224429, kim_antiferromagnetic_2021}; subsequent experiments--thanks to the availability of single crystals--proposed an alternative scenario with a noncollinear multi-$Q$ ground state\,\cite{li-22prx041024,kruger-23prl146702,jin-25prl136701,bischof-25prb104406,bestha-26prb100412}.

In contrast to the large number of antiferromagnetically ordered honeycomb magnets, comparatively little attention has been paid to ferromagnetic Kitaev materials. Some primary candidate materials in this category include $\text{CrI}_3$\,\cite{PhysRevLett.124.017201, CrI3_2, chen_magnetic_2020, CrBr3_CrI3_CrCl3} and $\text{CrBr}_3$\,\cite{CrBr3_CrI3_CrCl3}. While ferromagnets have been studied extensively, most are adequately described by a Heisenberg-type model supplemented by easy-plane anisotropy. Here we investigate \LCSO\ in its monoclinic honeycomb phase\,\footnote{Note that $\text{Li}_{3}\text{Co}_2 \text{SbO}_6$ can be obtained in two distinct polymorphs depending on synthetic conditions \cite{Synthesis_for_LCSO}: (i) the monoclinic honeycomb phase, of interest here, and (ii) an orthorhombic $Fddd$ phase.}, stoichiometrically equivalent to \NCSO, by means of INS experiments. It was previously shown to possess ferromagnetically ordered planes, where the magnetization direction alternates between adjacent layers\,\cite{Synthesis_for_LCSO, stratan_synthesis_2019, PhysRevB.102.224411} (referred to as an ``A-type'' antiferromagnet in the literature). This makes \LCSO\ particularly intriguing, especially in light of its sister compound \NCSO, which is believed to exhibit substantial Kitaev interactions. In this work, we have collected INS data for zero field, but also for finite magnetic fields up to 7\,T. The experimental measurements are complemented by a detailed analysis using linear spin wave theory (LSWT). By fitting the calculated dynamical structure factor (DSF) to the INS data, we find that a sizeable Kitaev interaction, together with symmetric off-diagonal $\Gamma$ and $\Gamma'$ exchanges, can achieve quantitative agreement with the experiment.

\begin{figure}[b!]
    \centering
    \includegraphics[width=0.85 \linewidth]{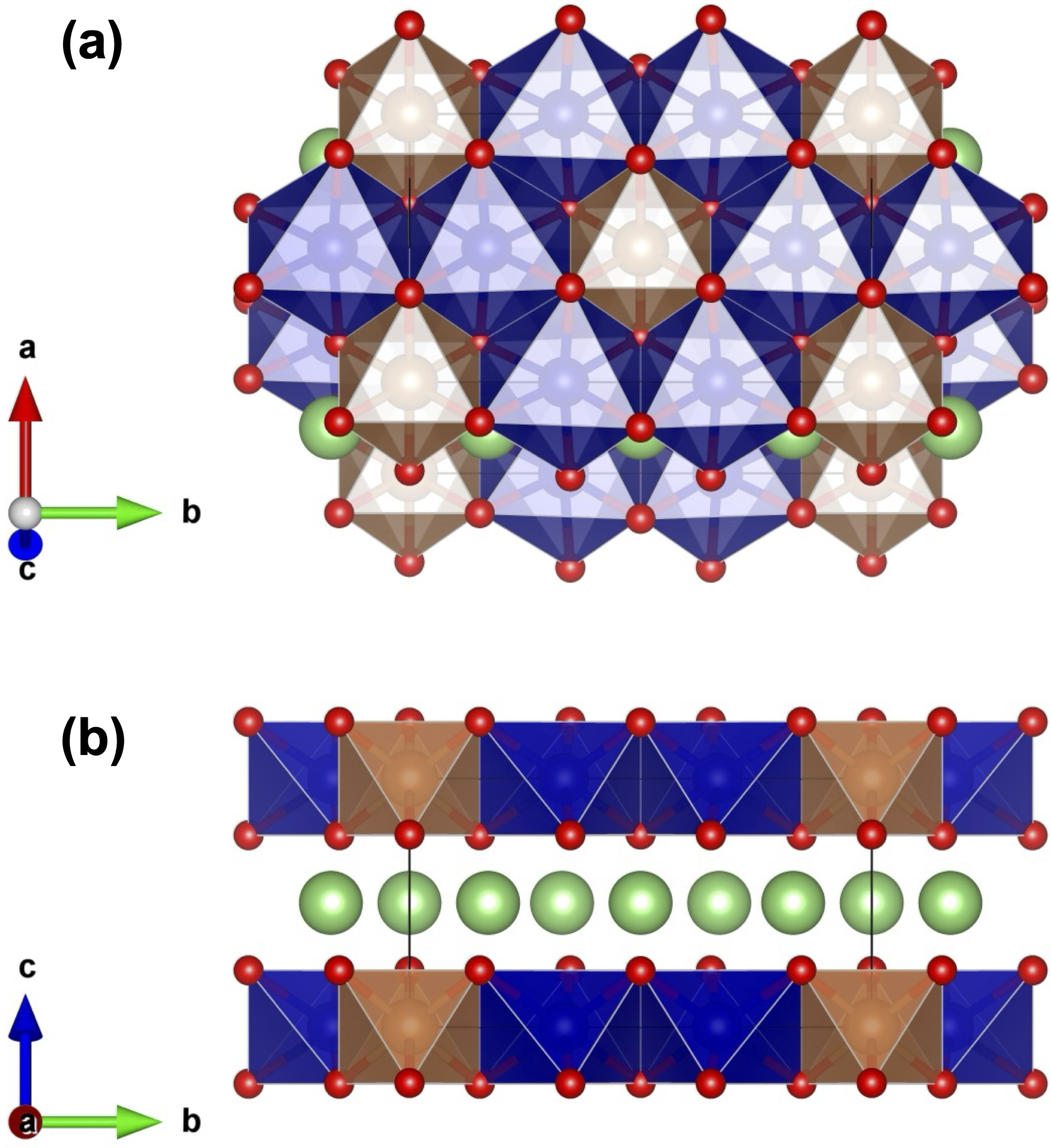}
    \caption{Crystal structure of the honeycomb phase of \LCSO\ viewed (a) perpendicular and (b) parallel to the honeycomb layers. CoO$_6$ octahedra are blue, SbO$_6$ octahedra are brown, Li atoms are green, and O atoms are red. Space group $C2/m$, 298 K lattice parameters $a=5.2717(6)$ \AA, $b=9.0871(10)$ \AA, $c=2.2039(6)$ \AA, $\beta=110.1553(14)$\textdegree.}
\label{fig:LCSO_Honeycomb_phase}
\end{figure}

Sec.\,\ref{sec:sample-prep} describes the sample preparation and crystal structure. The INS measurements and theoretical modelling are presented in Secs.\,\ref{sec:INS} and \ref{sec:SWT}, respectively. Sec.\,\ref{sec:in-field} reports the magnetic-field–dependent response up to 7 T, and Sec.\,\ref{sec:conclusion} provides our conclusions. 

%
%
\section{Sample preparation}
\label{sec:sample-prep}

A $\sim$6 g sample of the monoclinic phase of \LCSO\ was synthesized by the direct solid-state method reported previously \cite{Synthesis_for_LCSO}. The sample was confirmed to be pure by Rietveld-refinement against X-ray powder diffraction data. The crystal structure is shown in Fig.\,\ref{fig:LCSO_Honeycomb_phase}, highlighting the honeycomb arrangement of Co due to the strict 2:1 ordering of CoO$_6$ and SbO$_6$ edge-sharing octahedra. The honeycomb layers are separated by layers of Li in octahedral interstitial sites (O3-type stacking arrangement).

%
%
\section{Inelastic Neutron Scattering}
\label{sec:INS}
INS data were collected using the cold-neutron time-of-flight spectrometer Pelican\,\cite{pelican1,pelican2} at the Australian Centre for Neutron Scattering.  The sample was held in an annular aluminium sample can 
inserted inside a top-loading cryomagnet.  The instrument was optimised for $\lambda = 4.69\ \text{\AA}$ neutrons.  The choppers were also phased to allow the collection of data with $\lambda/2 = 2.345\ \text{\AA}$ neutrons. Data were collected at 1.6 K, 9 K and 20 K with both wavelengths.  Additional data were collected at 0.5, 1, 2, 3, 5 and 7 T with a sample temperature of 1.6 K.

\begin{figure}[t!]
    \centering
    \includegraphics[width=1 \linewidth]{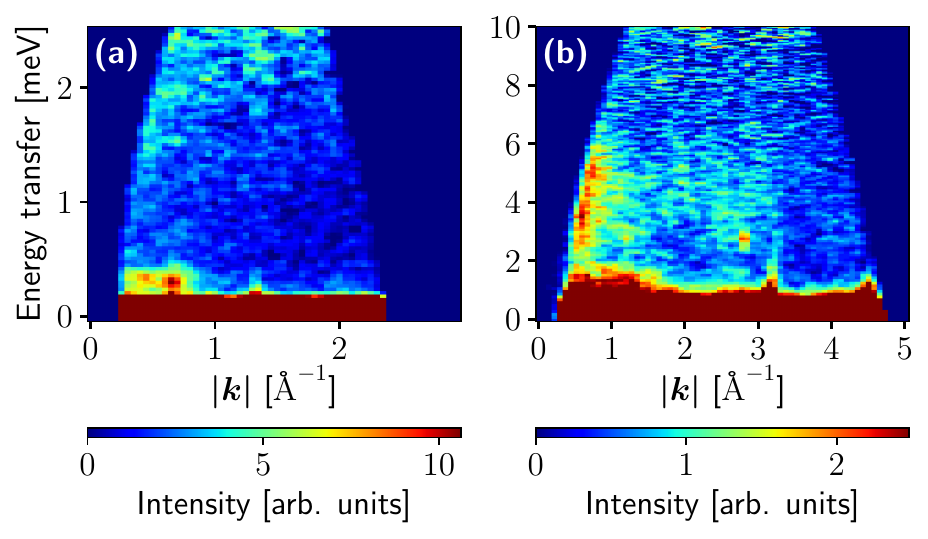}
    \caption{Powder-averaged INS spectra measured using neutrons with wavelengths of (a) $\lambda = 4.69\,\text{\AA}$ and (b) $\lambda = 2.345\,\text{\AA}$. The sample temperature for these measurements was set to 1.6\,K.} 
    \label{fig: INS_0T}
\end{figure}

Magnetic Bragg reflections were observed below 15 K at  $|\vec{k}| = 0.64\,\text{\AA}^{-1}$ confirming that the system is magnetically ordered and the previously published magnetic ordering transition of $T_N = 14$ K.  The spectrum of \LCSO\ measured with $\lambda/2 = 2.345\, \text{\AA}$ neutrons at 1.6 K shows a broad excitation at the low $|\vec{k}|$ kinematically allowed limit between $\approx$ 2 meV and $\approx$ 6 meV.  Measurements at 20 K with both $\lambda = 4.69\, \text{\AA}$ and $\lambda/2 = 2.345\, \text{\AA}$ neutrons show a broad quasi elastic contribution which decreases with increasing $|\vec{k}|$. This is consistent with the low temperature data having magnon excitations emanate from the $\Gamma$ point, whereas above $T_N$ there are spin fluctuations and short range magnetic order. The zero-field INS spectra measured using $\lambda = 4.69\, \text{\AA}$ and $\lambda/2 = 2.345\ \text{\AA}$ neutrons are presented in Figs.\,\hyperref[fig: INS_0T]{\ref*{fig: INS_0T}(a)} and \hyperref[fig: INS_0T]{\ref*{fig: INS_0T}(b)}, respectively.  

%
%

%
%
\section{Spin Wave Theory}
\label{sec:SWT}
As for the Hamiltonian modeling \LCSO, we consider the extended Kitaev-Heisenberg spin model. It includes, within each honeycomb plane, the bond-dependent anisotropic Kitaev term ($K$), the nearest neighbor isotropic Heisenberg term ($J_1$) and the nearest neighbor off-diagonal spin coupling terms (${\Gamma}, {\Gamma'}$). In addition, we also include a Heisenberg term ($J_\perp$) that couples neighboring spins in adjacent planes. The total Hamiltonian reads
\begin{align}
\mathcal{H} &= \sum_{\braket{i j}^\gamma} J_1 \boldsymbol{S}_i \cdot \boldsymbol{S}_j + K S^{\gamma}_i S^{\gamma}_j + \Gamma \left(S^{\alpha}_i S^{\beta}_j + S^{\beta}_i S^{\alpha}_j\right) \nonumber\\[5pt]
&+ \Gamma' \left(S^{\gamma}_i S^{\alpha}_j + S^{\alpha}_i S^{\gamma}_j + S^{\gamma}_i S^{\beta}_j + S^{\beta}_i S^{\gamma}_j \right) \nonumber\\[10pt]
&+ \sum_{\braket{ij}_{\perp}} J_{\perp} \boldsymbol{S}_i \cdot \boldsymbol{S}_j\ .
\label{eqn: Extended_KH_Hamiltonian}
\end{align}
The indices $\{\alpha, \beta, \gamma \}$ in Eq.\eqref{eqn: Extended_KH_Hamiltonian} of the spin components are ordered as cyclic permutations of $\{x, y, z \}$. The term $\sum_{\braket{i j}^\gamma}$ corresponds to the sum of all nearest-neighboring pairs of lattice sites $(i,j)$ categorized by their Ising-type bond link, $\gamma$. The second sum, $\sum_{\braket{ij}_\perp}$, corresponds to all neighboring inter-plane pairs of lattice sites $(i,j)$. The matrix representation of the Hamiltonian terms within each plane is given by
\begin{align}
    \mathcal{H}_{ij}^{(\gamma)} = \vec{S}_i^{T} {H}_\gamma \vec{S}_j\ ,
    \label{eqn: extended_KH_matrix}
\end{align}
with the three spin coupling matrices
$$
H_x = \begin{pmatrix}
    K + J_1 & \Gamma' & \Gamma' \\
    \Gamma' & J_1 & \Gamma \\
    \Gamma' & \Gamma & J_1
\end{pmatrix}, \hspace{0.5em}
    H_y = \begin{pmatrix}
    J_1 & \Gamma' & \Gamma \\
    \Gamma' & K + J_1 & \Gamma' \\
    \Gamma & \Gamma' & J_1
\end{pmatrix},
$$
\begin{align}
    H_z = \begin{pmatrix}
        J_1 & \Gamma & \Gamma' \\
        \Gamma & J_1 & \Gamma' \\
        \Gamma' & \Gamma' & K +J_1
        \end{pmatrix}\ .
    \label{eqn: K_matrices}
\end{align}
The neutron powder diffraction measurements reported in Refs.\,\cite{Synthesis_for_LCSO} and \cite{stratan_synthesis_2019}, along with the measurements presented in Sec.\,\ref{sec:in-field}, indicate that the system hosts an A-type antiferromagnetic ground state, a configuration in which spins align ferromagnetically within each honeycomb layer and are antiferromagnetically coupled between adjacent layers. This magnetic stacking corresponds to a positive interlayer exchange term, $J_\perp>0$. Due to the lack of a global SU(2) symmetry in the model, the spin quantization axes are fixed with respect to the crystallographic planes. The microscopic model sets them perpendicular to their corresponding Kitaev links\,\cite{chaloupka_hidden_2015, Jackelli_Khaliulin_khomskii}, i.e, $\hat{S}^x$ would be perpendicular to $x$-bonds while $\hat{S}^y$ and $\hat{S}^z$ would be perpendicular to $y$-bonds and $z$-bonds, respectively. As a result, the magnetization direction with respect to this new spin basis has to be determined first before considering quantum fluctuations around it.

\subsection{Classical Ground State configuration}
\label{sec:Classic_GS}
Strategies to determine the true classical ground state configuration for frustrated magnetic systems include the Lutinger-Tizsa method\,\cite{Luttinger_Tisza_1, Luttinger_Tisza_2}, classical Monte Carlo simulations\,\cite{canting_2, spin_flop_1} and exact diagonalization\,\cite{Na2IrO3}; in the presence of a magnetic field, one is limited to the latter two methods.
From the neutron powder diffraction data we can assume a ferromagnetic ground state order within the honeycomb planes. With this input, the classical Hamiltonian would then be minimized with respect to the magnetization direction, accomplished by diagonalization of the matrix $H_{cl}=H_x+H_y+H_z$,
\begin{equation}
    H_{cl} = \begin{pmatrix}
        3J_1 + K & \Gamma + 2\Gamma' & \Gamma + 2\Gamma' \\[5pt]
        \Gamma + 2\Gamma' & 3J_1 + K & \Gamma + 2\Gamma' \\[5pt]
        \Gamma + 2\Gamma' & \Gamma + 2\Gamma' & 3J_1 + K
    \end{pmatrix}\ .
    \label{eqn: classical_matrix}
\end{equation}
For a detailed explanation, see Appendix\,\ref{sec:variational_minimization}. One can neglect the diagonal entries in Eq.\,\eqref{eqn: classical_matrix} since they correspond to a constant shift in the eigenvalues. The resulting eigenvectors $\vec{v_i}$ and their corresponding eigenvalues $\lambda_i$ are thus independent of $J_1$ and given by
\begin{equation}
    \vec{v}_1 = (1, 1, 1), \hspace{0.5em}
\vec{v}_2 = (-1, 1, 0), \hspace{0.5em}\vec{v}_3 = (-1, -1, 2),
    \label{eqn: eigenvectors_ferromagnetic}
\end{equation}
and
\begin{equation}
    \lambda_1 = 2(\Gamma + 2\Gamma'),
\hspace{0.5em} \lambda_2 = \lambda_3 = - (\Gamma + 2\Gamma').
    \label{eqn: eigenvalues_ferromagnetic}
\end{equation}
The magnetization direction is then determined by choosing the eigenvector corresponding to the smallest eigenvalue. From Eq.\,\eqref{eqn: eigenvalues_ferromagnetic} we can see that the classical orientation of spins depends on the sign of the expression $\Gamma + 2\Gamma'$. Specifically, if $\Gamma + 2\Gamma' < 0$, the system favors an out-of-plane magnetization, i.e., $\vec{v}_1$ is chosen. On the other hand, if $\Gamma + 2\Gamma' > 0$, the magnetization direction falls within the honeycomb plane with each direction being energetically degenerate on a classical level, i.e., any unit vector formed by a linear combination of $\vec{v}_2$ and $\vec{v}_3$ is equally favorable. In our model, the inclusion of second-order quantum corrections (zero-point energies) would lift this degeneracy and favor a few in-plane directions which are symmetry-related. Previous neutron powder diffraction studies\,\cite{stratan_synthesis_2019, Synthesis_for_LCSO} revealed that for \LCSO, the ordered ground state has an in-plane spin configuration. 

\subsection{Bosonic Diagonalization}
\label{sec:Bosonic_Diag}
Once the ground state configuration has been determined, the system would then have to be redescribed using a new spin basis. This new spin basis should have the $z$-component directed parallel to the local classical moment. This kind of spin alignment transformation can be achieved using local SO(3) rotation matrices. For a given classical spin configuration, however, these SO(3) rotation matrices are not unique. There is a local {U(1)} gauge degree of freedom associated with additional transverse rotations along the classical moment\,\cite{Magnon_damping}. These, however, leave the excitation spectrum unchanged.

To employ spin wave theory, we express the transformed Hamiltonian using Holstein-Primakoff spin bosons\,\cite{Enrico_Rastelli}. The transformation is defined as
\begin{align}
    S_{i, \lambda}^{z} &= S - a_i^{(\lambda)\dagger} a_i^{(\lambda)},  \nonumber \\
    S_{i, \lambda}^{+} &= \sqrt{2S} \sqrt{1 - \frac{a_i^{(\lambda)\dagger} a_i^{(\lambda)}}{2S}} \hspace{0.5 em}a_i^{(\lambda)}, \nonumber\\
    S_{i, \lambda}^{-} &= \sqrt{2S} a_i^{(\lambda)\dagger} \hspace{0.5 em} \sqrt{1 - \frac{a_i^{(\lambda)\dagger} a_i^{(\lambda)}}{2S}}\ ,
    \label{eqn: Holstein_primakoff}
\end{align}
where $S_{i, \lambda}^{\pm}$ represents the spin raising (lowering) operators at unit cell, $i$, and  sublattice, $\lambda$.

LSWT assumes a dilute magnon population at low temperatures, allowing magnon–magnon interactions to be neglected and reducing the Hamiltonian to its quadratic (non-interacting) form. Practically, this amounts to truncating the Holstein-Primakoff expansion in Eq.\,\eqref{eqn: Holstein_primakoff} to leading order in 
$1/S$. The classical configuration corresponds to the zeroth-order term ($S \rightarrow \infty$), while LSWT incorporates the first-order quantum corrections through free-magnon excitations. The presence of a true classical ground state configuration is further reinforced by the absence of any linear bosonic terms in the transformed Hamiltonian. The presence of any such term would imply that the ground state configuration is unstable. For the given magnetic ordering of \LCSO, a 4-site magnetic unit cell has been chosen (2 sites for each plane). With the virtue of Fourier transforms, a bosonic quadratic Hamiltonian can be constructed as follows,
\begin{align}
    \mathcal{H} = \frac{1}{2}\sum_{\vec{k}} \Psi_{\vec{k}}^\dagger \,H_{\vec{k}}^{(2)}\, \Psi_{\vec{k}}\ .
    \label{eqn: Bdg_hamiltonian}
\end{align}
The 8-dimensional vector $\Psi_{\vec{k}}$ consists of bosonic creation and annihilation operators for different sublattices defined within the magnetic unit cell:
\begin{align}
    \Psi_{\vec{k}} = \begin{pmatrix}
    a_{\vec{k}}^{(1)} & a_{\vec{k}}^{(2)} & a_{\vec{k}}^{(3)} & a_{\vec{k}}^{(4)} & a_{-\vec{k}}^{(1)\dagger} & a_{-\vec{k}}^{(2)\dagger} & a_{-\vec{k}}^{(3)\dagger} & a_{-\vec{k}}^{(4)\dagger}  \end{pmatrix}^{T}.
    \label{eqn: BDG_basis}
\end{align}
The 8-by-8 matrix ${H_{\vec{k}}^{(2)}}$ is generally not diagonal. To find the magnon spectra, this matrix has to be diagonalized in a way that preserves the bosonic commutation relations. This can be achieved using a Bogoliubov transformation, ${{B}_{\vec{k}}} \Psi_{\vec{k}} = \Phi_{\vec{k}}$, that diagonalizes ${H_{\vec{k}}^{(2)}}$ and satisfies the following relations,
\begin{equation}
    {B}_{\vec{k}} {G} {B}_{\vec{k}}^\dagger = {G}\hspace{0.5 em}\Leftrightarrow  \hspace{0.5 em} {B}_{\vec{k}}^\dagger {G} {B}_{\vec{k}} = {G}
    \label{eqn: Bogoliubov_trans_commutation}
\end{equation}
and 
\begin{equation}
    P B_{\vec{k}} P = B_{-\vec{k}}^{*}.
    \label{eqn: self_consistency}
\end{equation}
The commutator matrix ${G}$ and the permutation matrix $P$ are defined as
\begin{equation}
    {G} = \begin{pmatrix}
        \mathbb{1}_{4\times4} & 0 \\
        0 & -\mathbb{1}_{4\times4}
    \end{pmatrix}
    \label{eqn: commutator_matrix}
\end{equation}
and 
\begin{equation}
    P = \begin{pmatrix}
        0 & \mathbb{1}_{4\times4} \\
        \mathbb{1}_{4\times4} & 0
    \end{pmatrix}.
    \label{eqn: permutation matrix}
\end{equation}
The new vector $\Phi_{\vec{k}}$ is given by
\begin{align}
\Phi_{\vec{k}} = \begin{pmatrix}
    \alpha_{\vec{k}}^{(1)} \!& \alpha_{\vec{k}}^{(2)} \!& \alpha_{\vec{k}}^{(3)} \!& \alpha_{\vec{k}}^{(4)} \!& \alpha_{-\vec{k}}^{(1)\dagger} \!& \alpha_{-\vec{k}}^{(2)\dagger} \!& \alpha_{-\vec{k}}^{(3)\dagger} \!& \alpha_{-\vec{k}}^{(4)\dagger}
    \end{pmatrix}^{T},
    \label{eqn: Magnon_modes_vector}
\end{align}
where $\alpha_{\vec{k}}^{(n)\dagger}$ and $\alpha_{\vec{k}}^{(n)}$ correspond to magnon creation and annihilation operators, respectively.
Here, ${B}_{\vec{k}}$ is not a unitary matrix and thus requires careful handling. Fortunately, Colpa's algorithm\,\cite{Colpa_Algorithm} provides a systematic way of diagonalizing $H_{\vec{k}}^{(2)}$ using Cholesky decomposition.

\subsection{Dynamical Structure Factor}
\label{sec:DSF}
With the magnon excitation spectra evaluated, we now need to find their intensities. The differential cross-section for neutron scattering is given by
\begin{equation}
    \frac{d^2 \sigma}{d\Omega dE_f} \propto F(\vec{k})^2 S_{\perp}(\vec{k}, \omega),
    \label{eqn: diff_cross_section}
\end{equation}
with $F(\vec{k})$ being the atomic form factor for the $\text{Co}^{2+}$ ions \cite{Co2+_form_factor} and $S_{\perp}(\vec{k}, \omega)$ being the DSF defined as 
\begin{equation}
    S_{\perp}(\vec{k}, \omega) = \sum_{\alpha \beta} \left(\delta_{\alpha \beta} - \frac{k^\alpha k^\beta}{|\vec{k}|^2} \right) S^{\alpha \beta}(\vec{k}, \omega).
    \label{eqn: DSF}
\end{equation}
The matrix elements of the DSF are given by
\begin{equation}
    S^{\alpha \beta}(\vec{k}, \omega) = \sum_{i,j} \int_{-\infty}^\infty \frac{d\tau}{2\pi} e^{-i\omega \tau} \braket{S_{-\vec{k}i}^{\alpha}(0) S_{\vec{k}j}^{\beta}(\tau)},
    \label{eqn: DSF_elements}
\end{equation}
where the indices $(i,j)$ correspond to inequivalent lattice sites within the magnetic unit cell.
$S_{\vec{k}}^\alpha (t)$ represents the Fourier transform of the spin operator $S_i^\alpha$ in the Heisenberg picture. In the linear spin wave description, the time dependence of the spin operator is propagated to the bosonic creation and annihilation terms via
\begin{equation}
    \alpha_{\vec{k}}^{(i)}(t) = e^{-i\hbar \omega_{\vec{k}}^{(i)} t} \alpha_{\vec{k}}^{(i)}. 
\end{equation}
The INS spectrum is obtained from a powdered sample of $\text{Li}_{3}\text{Co}_2 \text{SbO}_6$. To keep this accounted for in our theoretical description, the DSF \eqref{eqn: DSF} has to be averaged over the unit sphere in $k$-space as
\begin{equation}   
    S_{\rm avg}(|{\vec{k}}|, \omega) = \frac{1}{4 \pi} \int d\Omega \hspace{0.25 em}S_{\perp}({\vec{k}}, \omega).
    \label{eqn: powdered_DSF}
\end{equation}
For a more detailed description of how the DSF is evaluated, see Appendix\,\ref{sec:appA}. 

\subsection{Comparison of INS data with LSWT}
\label{sec:comparison}
\begin{figure}[t!]
    \centering
    \includegraphics[width=0.94 \linewidth]{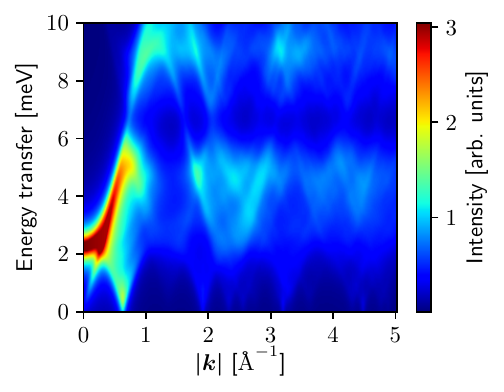}
    \caption{The DSF calculated for the best-fitted set of model parameters. The chosen parameter set is $(J_1, K, \Gamma, \Gamma', J_{\perp})=(-1.57, -5.90, 0.97, 0.35, 0.50)$ meV. A Lorentzian broadening of $\gamma = 0.45$ meV was chosen phenomenologically.} 
    \label{fig: DSF_fitted}
\end{figure}

To provide a quantitative picture for the comparison, the best-fitted parameters are chosen based on the minimization of the term
\begin{equation}
     \chi = \sum_{|\vec{k}|, \omega} \frac{\mathcal{I}_{\text{experimental}} - \mathcal{I}_{\text{theoretical}}}{\mathcal{I}_{\text{theoretical}}},
    \label{eqn: Chi_stat}
\end{equation}
where we have chosen Fig.\,\hyperref[fig: INS_0T]{\ref*{fig: INS_0T}(b)} as the reference INS dataset, $\mathcal{I}_\text{experimental}$. Due to the limitations imposed by the experiment, the fitting parameter, $\chi$, would only consider points within the accessible window (allowed regions in the $|\vec{k}|-\omega$ space). Additionally, the broadened region near the elastic limit ($\omega \lesssim 2 \text{meV}$) masks any inelastic contributions and hence is neglected when fitting. The calculated DSFs have been renormalized in such a way that the total calculated intensity across the measured $|\vec{k}|-\omega$ space is set to be the same as the total intensity from the experimental INS spectrum. The best-fitted powder averaged DSF is shown in Fig.\,\ref{fig: DSF_fitted}. The region corresponding to $0.5\,\text{\AA}^{-1} \leq |\vec{k}| \leq 1\,\text{\AA}^{-1}$ and $2.5\,\text{meV} \leq E \leq 5\,\text{meV}$ shows a good overlap with the experiment. In particular, the DSF matches well with the band-like feature observed in the experiment.
Due to the lack of prominent high energy features in Fig.\,\hyperref[fig: INS_0T]{\ref*{fig: INS_0T}(a)}, we have not directly compared the DSF results to the INS data for $\lambda = 4.69\,\text{\AA}$, but we have qualitatively assessed the high intensity region found at $|\vec{k}| \approx 0.65\,\text{\AA}^{-1}$. Our analysis of the DSF indicates that there are soft modes ($\hbar \omega_{\vec{k}} = 0$) in the model. The emergence of soft modes, such as at $|\vec{k}| \approx 0$ and at $|\vec{k}| \approx 0.65\,\text{\AA}^{-1}$, is a result of the inherent U(1) degeneracy of the classical ground state spin configuration. As mentioned in Sec.\,\ref{sec:Classic_GS}, the in-plane orientation of spins remains continuously degenerate within the zeroth order of LSWT. This symmetry is subsequently lifted through spontaneous symmetry breaking and as a result, zero-energy excitations in the magnon spectrum emerge as Goldstone modes. This symmetry, however, is not present in the original quantum spin Hamiltonian and is usually referred to as an accidental symmetry \cite{Pseudo_goldstone_1, pseudo_goldstone_2}. As a consequence of this, the soft magnon modes are classified as pseudo-Goldstone modes instead. The high intensity feature in Fig.\,\hyperref[fig: INS_0T]{\ref*{fig: INS_0T}(a)} at $|\vec{k}| \approx 0.65 \mathrm{\AA}^{-1}$ is suspected to emanate from the soft mode corresponding to the antiferromagnetic interlayer ordering vector. Assuming the model does not host any hidden symmetry \cite{chaloupka_hidden_2015}, higher-order corrections in spin wave theory should open a gap in the spectrum \cite{Pseudo_goldstone_1, Na2IrO3, pseudo_goldstone_2, PRX_goldstone}.  
\\

For $\text{Li}_3\text{Co}_2\text{SbO}_6$, prior work has centered largely on its structural characterization using neutron powder diffraction and thermodynamic standard characterization\,\cite{stratan_synthesis_2019, Synthesis_for_LCSO}, without exploration of its magnetic and dynamical responses such as INS. Accordingly, the coupling parameters obtained here should be taken as informed estimates rather than precise values. The behavior of the DSF with respect to the model parameters is complex and interdependent, rendering any extrapolation technique difficult. Furthermore, there exists a duality transformation between parameter sets that emerges as a consequence of rotating the spin basis by $180^\circ$ \cite{Kitaev_materials_review_art, chaloupka_hidden_2015, Alaric_etal_new}. The self-duality transformation is described as
\renewcommand{\arraystretch}{1.5}
\begin{equation}
    \begin{pmatrix}
        \tilde{J}_1 \\
        \tilde{K} \\
        \tilde{\Gamma} \\
        \tilde{\Gamma'}
    \end{pmatrix} = \begin{pmatrix}
        1 & \frac{4}{9} & -\frac{4}{9} & \frac{4}{9}\\
        0 & -\frac{1}{3} & \frac{4}{3} & -\frac{4}{3}\\
        0 & \frac{4}{9} & \frac{5}{9} & \frac{4}{9} \\
        0 & -\frac{2}{9} & \frac{2}{9} & \frac{7}{9}
    \end{pmatrix} \begin{pmatrix}
        J_1 \\
        K \\
        \Gamma \\
        \Gamma'
    \end{pmatrix}.
    \label{eqn: Khaliulin_transformation}
\end{equation}
It implies that parameter sets that are equivalent under this self-duality transformation have the same powder-averaged DSF. 
Since this transformation emerges from the global rotation of spins, the interlayer Heisenberg coupling remains invariant.
\begin{table}[h]
    \centering
    \begin{tabular}{|l||c|c|c|c|c|}
     \hline
     \multicolumn{6}{|c|}{Model parameters (meV)} \\
     \hline
     Material & $J_1$ & $K$ & $\Gamma$ & $\Gamma'$ &$J_\perp$ \\
     \hline
     $\text{Li}_{3}\text{Co}_2 \text{SbO}_6$ (Set 1) & $-1.57$ & $-5.90$ & 0.97 & 0.35 & 0.50 \\
     \hline
     $\text{Li}_{3}\text{Co}_2 \text{SbO}_6$ (Set 2) & $-4.47$ & 2.79 & $-1.93$ & 1.80 & 0.50 \\
     \hline
    \end{tabular}
    \caption{Best-fitted model parameters along with their self-dual transformed counterparts.}
    \label{table: Khaliulin_coupling_params}
\end{table}
\\
\begin{figure}
    \centering
    \includegraphics[width=1\linewidth]{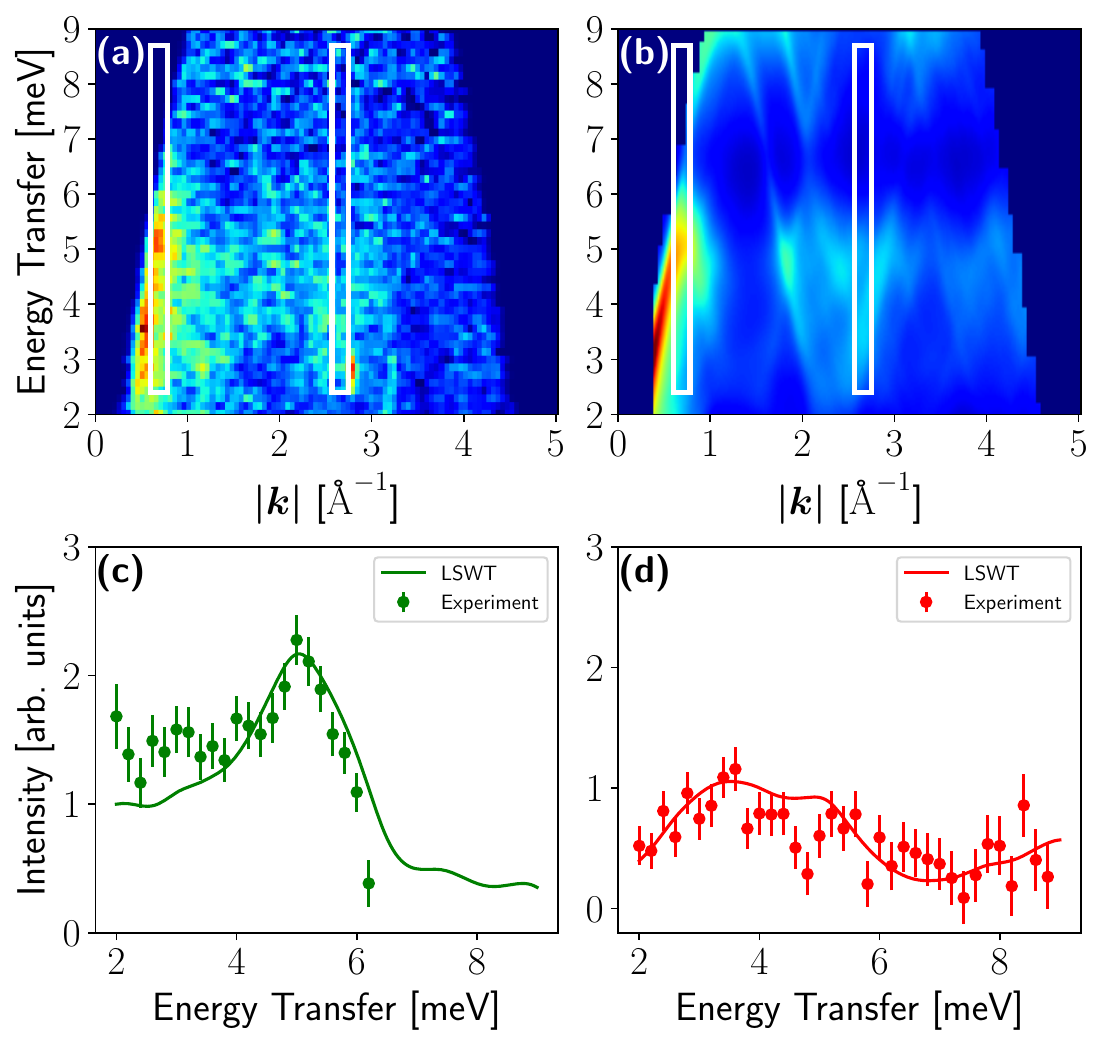}
    \caption{(a) Experimental INS spectrum used for fitting of the DSF. This is the same INS spectrum as shown in Fig.\,\hyperref[fig: INS_0T]{\ref*{fig: INS_0T}(b)} but without the elastically broadened region. (b) Theoretical, best-fitted powder-averaged DSF. The white stripes correspond to the constant-$k$ cuts taken for (c) $|\vec{k}| = 0.7 \pm 0.1 \,\text{\AA}^{-1}$ and (d) $|\vec{k}| = 2.65 \pm 0.1 \,\text{\AA}^{-1}$. (c-d) The continuous curves correspond to the results of LSWT while the discrete data points correspond to the experimental INS spectra. The model parameters from Table\,\ref{table: Khaliulin_coupling_params} have been used for all the theory figures.} 
    \label{fig: const_q_cuts}
\end{figure}
The best fit to the experimental INS spectrum in Fig.\,\hyperref[fig: const_q_cuts]{\ref{fig: const_q_cuts}(a)} is shown in Fig.\,\hyperref[fig: const_q_cuts]{\ref{fig: const_q_cuts}(b)}. Constant-$k$ cuts, indicated by the white stripes, at $|\vec{k}| = 0.7(0.1)\,\text{\AA}^{-1}$ and $|\vec{k}| = 2.65(0.1)\,\text{\AA}^{-1}$ are presented in Figs.\,\hyperref[fig: const_q_cuts]{\ref*{fig: const_q_cuts}(c)} and \hyperref[fig: const_q_cuts]{\ref*{fig: const_q_cuts}(d)}. The theoretical spectrum is calculated using the best-fitted parameters listed in Table.\,\ref{table: Khaliulin_coupling_params}. The fitting procedure was not restricted to the two constant-$k$ cuts shown in Figs.\,\hyperref[fig: const_q_cuts]{\ref*{fig: const_q_cuts}(a)} and \hyperref[fig: const_q_cuts]{\ref*{fig: const_q_cuts}(b)} but was carried out over the full detector window, excluding contributions arising from elastic broadening. There are also some unavoidable limitations arising from failing to capture spectral features that lie outside the accessible detector range. These features are, therefore, not constraint by the fit. Nevertheless, the overall agreement between the experimental data and the theoretical calculations are quantitatively reasonable.
%
%
\section{Results in a magnetic field}
\label{sec:in-field}
The response of frustrated magnetic systems to external magnetic fields has been shown to host a range of nontrivial phenomena. In this section, we investigate the effects of an externally applied magnetic field to the same powdered sample. We employ a combination of numerical energy minimization and LSWT to analyze the experimental observations obtained for \LCSO.

\subsection{Neutron powder diffraction}
\label{subsection: NPD_spectra}
Previous structural and thermodynamic investigations of Li$_3$Co$_2$SbO$_6$ have reported the presence of a metamagnetic transition, in which the A-type antiferromagnetic order is progressively suppressed and ultimately gives way to a fully polarized state\,\cite{Synthesis_for_LCSO,stratan_synthesis_2019}. In related theoretical work\,\cite{spin_flop_1}, it was demonstrated that spin-flop transitions may arise in several magnetically ordered phases of Kitaev systems, including stripy, zigzag, and Néel antiferromagnetic states while no intermediate phase has been predicted for the ferromagnetic regime\,\cite{spin_flop_1, Canting_isot}.

\begin{figure}[t!]
    \centering
    \includegraphics[width=1\linewidth]{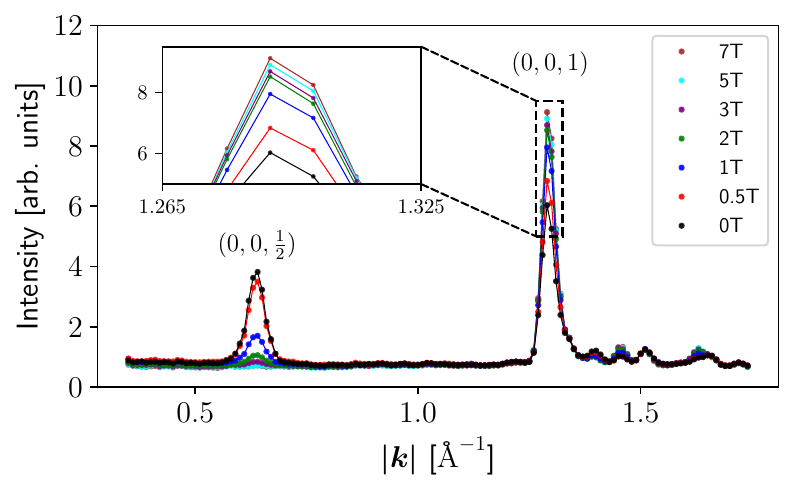}
    \caption{Neutron powder diffraction patterns in a range of applied magnetic fields. The magnetic Bragg reflection $(0,0,\tfrac{1}{2})$ is due to A-type antiferromagnetic order which is suppressed by the applied field, while the additional magnetic intensity on top of the nuclear Bragg reflection $(0,0,1)$ is due to a fully ferromagnetic state which is stabilized by the applied field.} 
    \label{fig:NPD_spectra}
\end{figure}

To investigate the field dependence of the magnetic order, we have extracted the neutron diffraction signal from the elastic line at a range of applied magnetic field strengths, as shown in Fig.\,\ref{fig:NPD_spectra}. 
The Bragg reflection $(0,0,\tfrac{1}{2})$ is characteristic of A-type antiferromagnetic ordering, which doubles the magnetic unit cell relative to the nuclear unit cell along the interlayer $c$ axis. Upon the application of an external magnetic field, the intensity of the $(0,0,\tfrac{1}{2})$ reflection decreases with increasing field to vanish at approximately 5\,T, consistent with the suppression of A-type order. Conversely, the intensity of the $(0,0,1)$ reflection increases with applied magnetic field, reflecting the progressive alignment of spins into a fully polarized ferromagnetic state where the magnetic unit cell is the same as the nuclear unit cell.

\subsection{INS spectra for external magnetic fields}
In the absence of $\Gamma$ and $\Gamma'$ terms in Eq.\,\eqref{eqn: eigenvalues_ferromagnetic}, the classical ground state exhibits an SO(3) degeneracy associated with arbitrary orientations of the magnetization. The inclusion of finite $\Gamma$ and $\Gamma'$ interactions with $\Gamma + 2 \Gamma' > 0$ reduces the SO(3) to a classical U(1) symmetry, remaining consistent with A-type antiferromagnetic order due to the isotropic nature of the interlayer coupling. At the classical level, the determination of the spin configuration in the presence of an external magnetic field is typically carried out using numerical optimization techniques, such as iterative energy minimization or Monte Carlo simulations. In the present work, we adopt a simplified approach by assuming that the chosen magnetic unit cell remains stable upon application of an external magnetic field. This assumption is motivated by the classical degeneracy associated with the in-plane magnetization direction, which effectively defines an easy plane in the presence of an infinitesimal Zeeman field.

\begin{figure}[t!]
    \centering
    \includegraphics[width=0.81 \linewidth]{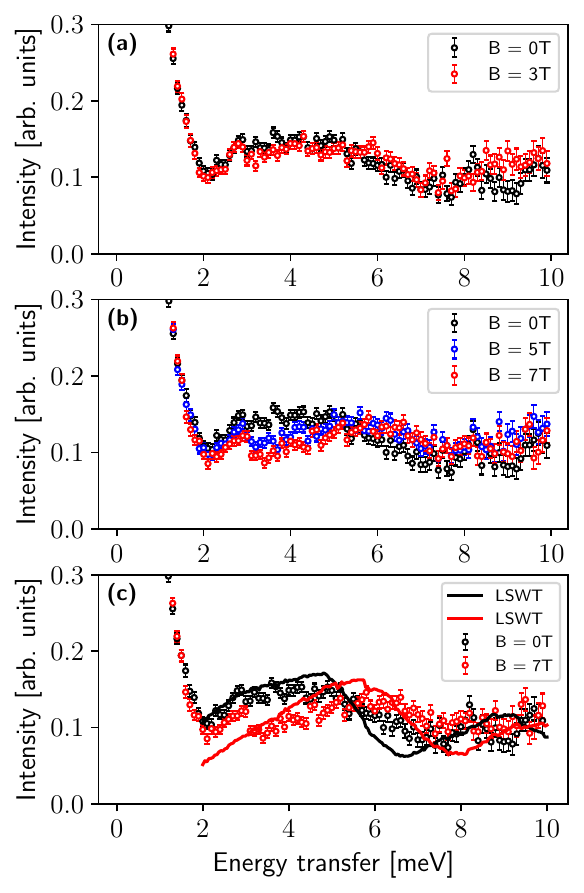}
    \caption{(a-b) Momentum-integrated INS spectra measured at different applied magnetic field strengths. (c) Momentum-integrated DSFs calculated for in-plane external magnetic fields (curves), overlaid on the corresponding experimental INS spectra (data points).} 
    \label{fig: B_field_overlay_combined}
\end{figure}

As the field strength increases, the classical spins are expected to cant toward the direction of the applied field. This description remains valid provided that the underlying in-plane ferromagnetic configuration is preserved. To substantiate this assumption, we perform a finite system-size numerical minimization calculation, which confirms the stability of the adopted magnetic unit cell. 

To this end, we aim to reproduce features of the INS spectra in an external field by adding a Zeeman term to our Hamiltonian \eqref{eqn: Extended_KH_Hamiltonian},
\begin{equation}
    H_{ext} = -\vec{h} \cdot \sum_{i} \vec{S}_i,
    \label{eqn: Zeeman_term}
\end{equation}
where $i$ labels all lattice sites and $\vec{h}$ is the external Zeeman field. The classical spin configurations are highly sensitive to the direction of the applied field, which complicates a direct theoretical treatment of the DSF for powdered samples. For different field orientations, the system stabilizes distinct classical spin configurations. Consequently, arbitrary orientations of crystal planes are not captured within our model by a simple averaging of the DSF over the unit sphere in $\vec{k}$-space alone. To properly model such samples under applied magnetic field, DSFs have to be computed for all-field directions and all $k$-directions,
\begin{equation}   
    S_{\rm avg}(|{\vec{k}}|, |\vec{h}|, \omega) = \frac{1}{(4 \pi)^2} \int \int d\Omega_{\vec{k}} d\Omega_{\vec{h}} \hspace{0.25 em}S_{\perp}(\vec{k}, \vec{h}, \omega),
    \label{eqn: powdered_DSF_mag_ext}
\end{equation}
where $S_{\perp}(\vec{k}, \vec{h}, \omega)$ is the DSF calculated for a given Zeeman field, $\vec{h}$ and a momentum transfer, $\vec{k}$. Such a treatment would, therefore, involve minimizing classical spin orientations and DSF implementation for every single field direction. This is computationally expensive and as such, the DSFs are computed for a set of representative field orientations instead. These results are then compared with the experimental INS data.

The INS spectra measured at $\lambda = 2.345\,\text{\AA}$ exhibit a gradual evolution with increasing external magnetic field strength. The momentum-integrated spectra (across the measured range) remain largely unchanged up to 3\,T, while more pronounced changes appear at higher fields, particularly for $\text{B} \geq 5$\,T, as shown in Figs.\,\hyperref[fig: B_field_overlay_combined]{\ref*{fig: B_field_overlay_combined}(a)} and \hyperref[fig: B_field_overlay_combined]{\ref*{fig: B_field_overlay_combined}(b)}. In this regime, a redistribution of spectral weight is observed, where the magnon bands seem to have been lifted slightly. This trend is further supported by comparison with the overlaid DSF (momentum-integrated) calculated for an in-plane magnetic field direction, shown in Fig.\,\hyperref[fig: B_field_overlay_combined]{\ref*{fig: B_field_overlay_combined}(c)}. Additional details on the field dependence and the role of different magnetic field orientations are provided in Appendices\,\ref{sec:appB} and \ref{sec:appC}.

\begin{figure}[t!]
    \centering
    \includegraphics[width=1 \linewidth]{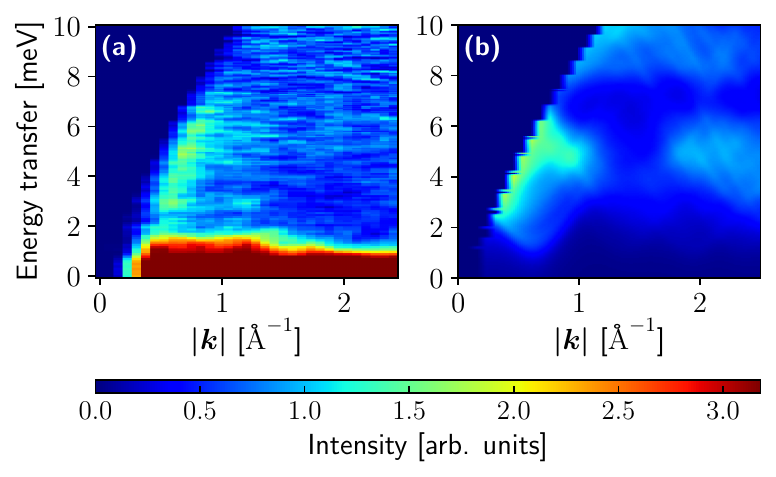}
    \caption{(a) INS spectrum measured at 1.6 K for an external magnetic field strength of 7\,T. (b) Powder averaged DSF for an in-plane magnetic field of same strength.} 
    \label{fig: 0vs7T_INS}
\end{figure}

In addition to the momentum-integrated data, the INS spectrum measured at a field strength of 7\,T is compared with the calculated DSF. As shown in Fig.\,\ref{fig: 0vs7T_INS}, a clear shift in spectral weight is observed in both the experimental and theoretical results. As discussed in Appendix\,\ref{sec:appC}, the spectral shift is most pronounced when the applied field is in-plane. In a powdered sample, however, the measured INS spectrum necessarily contains contributions from a distribution of field orientations. Nevertheless, as a qualitative benchmark, it is reasonable to expect that, apart from a limited set of special orientations (e.g., fields perpendicular to the honeycomb planes), the application of sufficiently strong fields generally induces a gap in the magnon spectrum as shown in Fig.\,\hyperref[fig: 0vs7T_INS]{\ref*{fig: 0vs7T_INS}(b)}.

%
%
\section{Conclusion}
\label{sec:conclusion}

In this work, we have investigated the magnetic properties of \LCSO\ using INS and neutron powder diffraction measurements. Within an extended Kitaev–Heisenberg spin model, the exchange coupling parameters were extracted by fitting the calculated DSF to the measured INS spectra. The resulting parameter set is characterized by a dominant ferromagnetic Kitaev exchange term, suggesting that \LCSO\ may represent a promising candidate for realizing Kitaev physics. While previous studies have reported neutron powder diffraction measurements alongside density functional theory and quantum-chemistry calculations, the present work provides the first INS-based investigation of this material. Notably, its sister compound, \NCSO, has been extensively studied via INS experiments; however, in contrast to \LCSO, its ground state is believed to stabilize a distinct magnetic order, namely a zig-zag phase or the more recently proposed triple-$Q$ state

In addition to the zero-field INS measurements, we performed experiments under applied magnetic fields of up to 7\,T. The corresponding neutron powder diffraction patterns reveal that, with increasing field strength, the Bragg peak associated with the interlayer antiferromagnetic order is progressively suppressed and vanishes near $\text{B} \approx 5\,\text{T}$, while the Bragg peak corresponding to the in-plane ferromagnetic component is simultaneously enhanced. This behavior is indicative of a transition toward a field-polarized or high-field phase. The INS spectra in this regime were subsequently compared with theoretical DSFs computed for the relevant classical ground-state configurations. We find good agreement of the INS data at B = 7\,T with LSWT calculations. Our work thus establishes \LCSO\, as a Kitaev A-type antiferromagnet with a non-negligible Kitaev exchange term.

\section*{Data availability}
The data that support the findings of this article are publicly available\,\cite{data}.

\acknowledgements
The authors acknowledge discussions with M.\ Vojta. 
S.R.\ acknowledges support from the Australian Research Council through Grant No.\ DP240100168. C.D.L.\ acknowledges support from the Australian Research Council through Grant No.\ DP230100558.
The authors acknowledge the support of the Australian Centre for Neutron Scattering, ANSTO and the Australian Government through the National Collaborative Research Infrastructure Strategy, in supporting the neutron research infrastructure used in this work via ACNS proposal DB13663. This research was supported by The University of Melbourne’s Research Computing Services and the Petascale Campus Initiative.



\appendix
\section{Details of the classical energy derivation}
\label{sec:variational_minimization}
The ferromagnetic ansatz in Sec.\,\ref{sec:Classic_GS} implies that every classical spin on a given honeycomb plane has the same orientation, $\vec{S}_i = \vec{v}$. As a result, the classical energy of the system can be written as
\begin{align}
    E_{cl} &= N_{uc}\frac{\vec{v}^{T} H_{cl} \vec{v}}{\vec{v}^{T} \vec{v}}
    = N_{uc}\frac{v^{i} H_{cl}^{ij} v^{j}}{v^{i} v^{i}}
    \label{eqn: classical_energy_matrix_spin}
\end{align}
where $N_{uc}$ is the total number of crystallographic unit cells in the system, $\vec{v} = (v^{x}, v^{y}, v^{z})^T$ and $H_{cl}$ is defined in Eq.\,\eqref{eqn: classical_matrix}. Note that by including the term $\vec{v}^T \vec{v}$ in Eq.\,\eqref{eqn: classical_energy_matrix_spin}, the orientation vector has been normalized. The overall scale of $\vec{v}$ does not affect the orientation of classical spins as long as the normalization constraint is imposed. Taking the derivative of \eqref{eqn: classical_energy_matrix_spin} with respect to the vector component $v^{k}$, we get
\begin{align}
    \frac{d E_{cl}}{dv^{k}} &= N_{uc}\frac{v^{i} H_{cl}^{ij} \delta_{jk} + H_{cl}^{ij} v^{j} \delta_{ik} - 2 E_{cl} v^{k}}{v^{i} v^{i}}
    \nonumber
    \\
    &= 2N_{uc}\frac{H_{cl}^{ki} v^{i} - E_{cl}v^{k}}{v^{i} v^{i}},
    \label{eqn: variational_derivation}
\end{align}
where repeated indices are summed over and the classical matrix is symmetric, $H_{cl}^{ij} = H_{cl}^{ji}$. Since Eq.\,\eqref{eqn: variational_derivation} holds for all $k$, by setting it equal to zero, we can reduce it to an eigenvalue problem given by
\begin{align}
    H_{cl} \vec{v} = E_{cl} \vec{v}\ .
    \label{eqn: EV_problem}
\end{align}
Note that this would give a stable solution if the given ansatz is true. For a general quadratic spin Hamiltonian, the Luttinger-Tisza method is often  employed. 

\section{Details of calculating the DSF}
\label{sec:appA}

To calculate the DSF \eqref{eqn: DSF}, one has to rewrite its components \eqref{eqn: DSF_elements} in the local spin-coordinate frame. This is achieved by using SO(3) rotation matrices that align the $z$-component of the locally transformed spins to the local classically minimized spins. This basis change transformation redescribes the classical ground state as a ferromagnet in the new frame, providing a convenient framework for the subsequent spin-flip excitations that eventually describe magnons. With this in mind, Eq.\,\eqref{eqn: DSF_elements} becomes
\begin{equation}
    S^{\alpha \beta}(\vec{k}, \omega) = \sum_{i,j} R_i^{\alpha \gamma} R_j^{\beta \xi} \int_{-\infty}^\infty \frac{d\tau}{2\pi} e^{-i\omega \tau} \braket{\tilde{S}_{-\vec{k}i}^{\gamma}(0) \tilde{S}_{\vec{k}j}^{\xi}(\tau)},
    \label{eqn: local_DSF_elements}
\end{equation}
where $\tilde{S}^{\gamma}$ represents the $\gamma$-component of the local frame spin and $R_i^{\alpha \gamma}$ corresponds to the $(\alpha \gamma)$ component of the matrix describing the local spin alignment transformation at the sublattice $i$. Note that the symbol $\braket{\cdot}$ represents a thermal average of all magnon states. This implies that terms of linear and cubic order vanish in the LSWT picture. Upon using the Holstein-Primakoff expansion and only keeping terms of the quadratic order, we can rewrite \eqref{eqn: local_DSF_elements} as
\begin{equation}
    S^{\alpha \beta} (\vec{k}, \omega) = \frac{S}{2}\int_{-\infty}^{\infty} \frac{d\tau}{2\pi} e^{-i\omega \tau} \braket{{\Psi}_{\vec{k}}^{\dagger}(0) {A}^{\alpha \beta} {\Psi}_{\vec{k}}(\tau)},
    \label{eqn: DSF_local_HP_mat}
\end{equation}
where ${A}^{\alpha \beta}$ is the overlap matrix for the component $(\alpha, \beta)$ and the spinor ${\Psi}_{\vec{k}} (\tau)$ is the time-evolved spinor in the Heisenberg picture defined in Eq.\,\eqref{eqn: BDG_basis}. Transforming these Holstein-Primakoff spinors into the relevant magnonic spinors, we get
\begin{equation}
    S^{\alpha \beta} (\vec{k}, \omega) = \frac{S}{2}\int_{-\infty}^{\infty} \frac{d\tau}{2\pi} e^{-i\omega \tau} \braket{{\Phi}_{\vec{k}}^{\dagger}(0) {O}_{\vec{k}}^{\alpha \beta} {\Phi}_{\vec{k}}(\tau)},
    \label{eqn:DSF_local_magnon_mat}
\end{equation}
where ${O}_{\vec{k}}^{\alpha \beta} = \left[({B}_{\vec{k}}^{-1})^\dagger {A}^{\alpha \beta} ({B}_{\vec{k}}^{-1})\right]$. Since the off-diagonal terms in ${O}_{\vec{k}}^{\alpha \beta}$ do not contribute to the thermal average, one would only consider terms that include the magnon number operators, i.e., 
\begin{equation}{
    \left\langle\alpha_{\vec{k}}^{(i)\dagger} \alpha_{\vec{k}}^{(i)}\right\rangle = \left\langle n_{\vec{k}}^{(i)} \right\rangle = \frac{1}{\exp{\left(\hbar \omega_{\vec{k}}^{(i)}/k_B T \right)} - 1}}\,
    \label{eqn: Bose_Einstein_magnon_1}
\end{equation}
and
\begin{equation}
    \left\langle\alpha_{\vec{k}}^{(i)} \alpha_{\vec{k}}^{(i)\dagger}\right\rangle = 1 + \left\langle n_{\vec{k}}^{(i)} \right\rangle\ ,
    \label{eqn: Bose_Einstein_magnon_1}
\end{equation}
as expected for non-interacting bosons. Finally, with all things considered, the DSF elements are written as
\begin{align}
    S^{\alpha \beta} (\vec{k}, \omega) &= \frac{S}{2} \sum_{l = 1}^n\Bigg[\delta\left(\omega + \omega_{\vec{k}}^{(l)}\right) \left\langle n_{\vec{k}}^{(l)}\right\rangle \left( {O}_{\vec{k}}^{\alpha \beta}\right)_{ll} \Bigg. \nonumber \\[5pt]
     &\Bigg.+ \delta\left(\omega - \omega_{\vec{k}}^{(l)}\right) \left(1 + \left\langle n_{\vec{k}}^{(l)}\right\rangle\right) \left( {O}_{\vec{k}}^{\alpha \beta}\right)_{(l + n)(l + n)}\Bigg],
    \label{eqn: DSF_final_inelastic_only}
\end{align}
where we have neglected terms including $\delta(\omega)$ since they correspond to elastic processes.

\section{Classical spin configuration in an external magnetic field}
\label{sec:appB}

%
\begin{figure}[t]
    \centering
    \includegraphics[width=0.9 \linewidth]{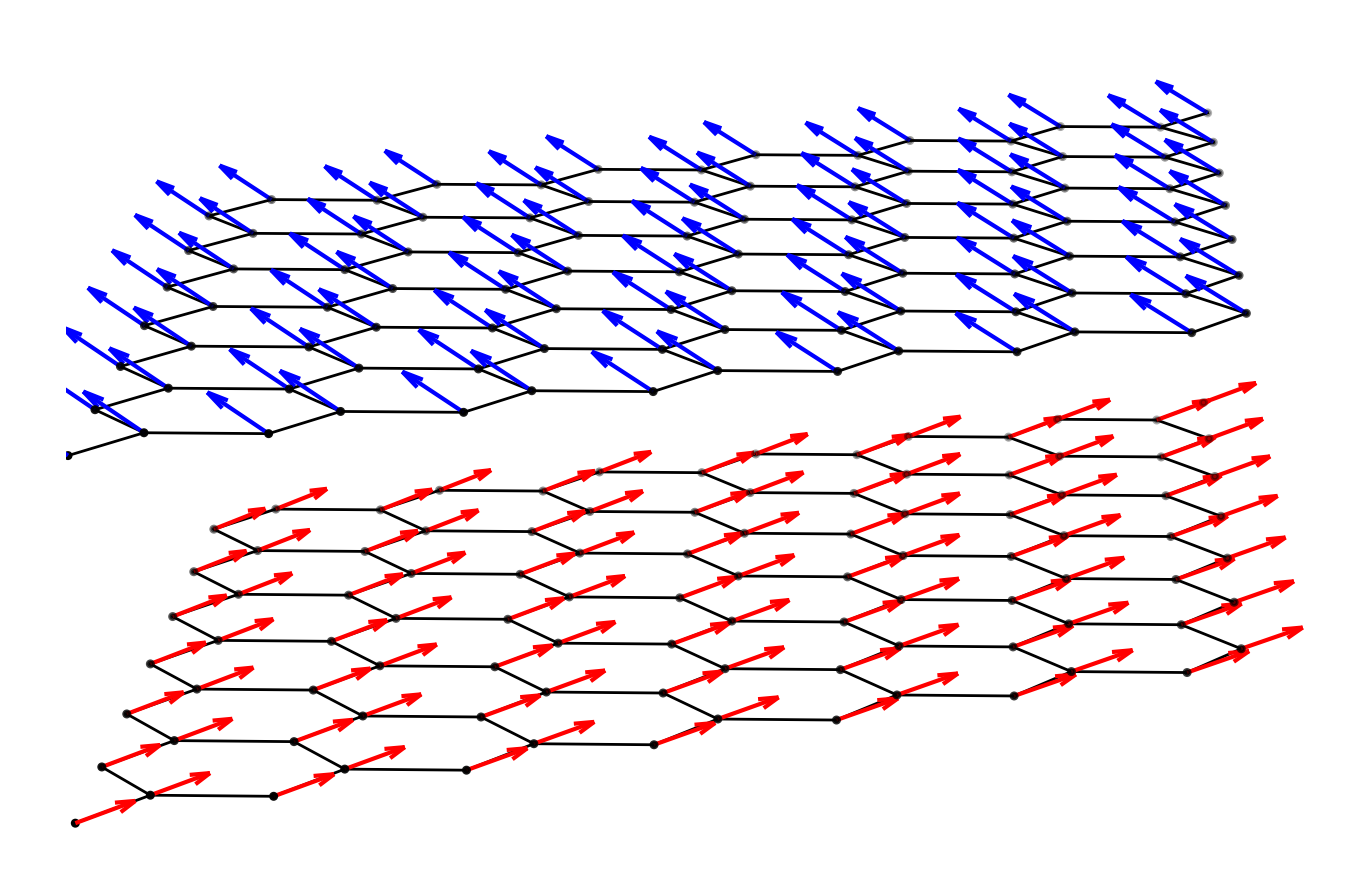}
    \caption{The classical ground state configuration in the presence of an out-of-plane external magnetic field. The figure shows ferromagnetic order within each plane and a homogenous canting induced within adjacent planes.} 
    \label{fig: GS_config_OOP}
\end{figure}
%

To determine the classical ground state configuration when the system is subject to an external Zeeman field, we have used a variational method that minimizes the classical energy assuming that the magnetic unit cell persists from the zero-field configuration. The system is iteratively minimized using 8 parameters corresponding to the direction of classical spins, i.e., two angular parameters for each lattice site. Fig.\,\ref{fig: GS_config_OOP} shows the classical spin configuration under an applied magnetic field directed perpendicular to the honeycomb planes. The induced spin canting between adjacent honeycomb layers is homogenous.
In addition to this, the classical magnetization of the ground state has been determined for different field orientations. The calculated expression is given by
\begin{equation}
    \vec{m}\cdot \vec{\hat{h}} = \frac{1}{4S}\sum_{i} \vec{S}_i \cdot \vec{\hat{h}}.
    \label{eqn: classical_mag_average}
\end{equation}
where the index $i$ corresponds to the four inequivalent lattice sites within the chosen magnetic unit cell and $\vec{S}_i$ is the local spin moment.

\begin{figure}[t!]
    \centering
    \includegraphics[width=1 \linewidth]{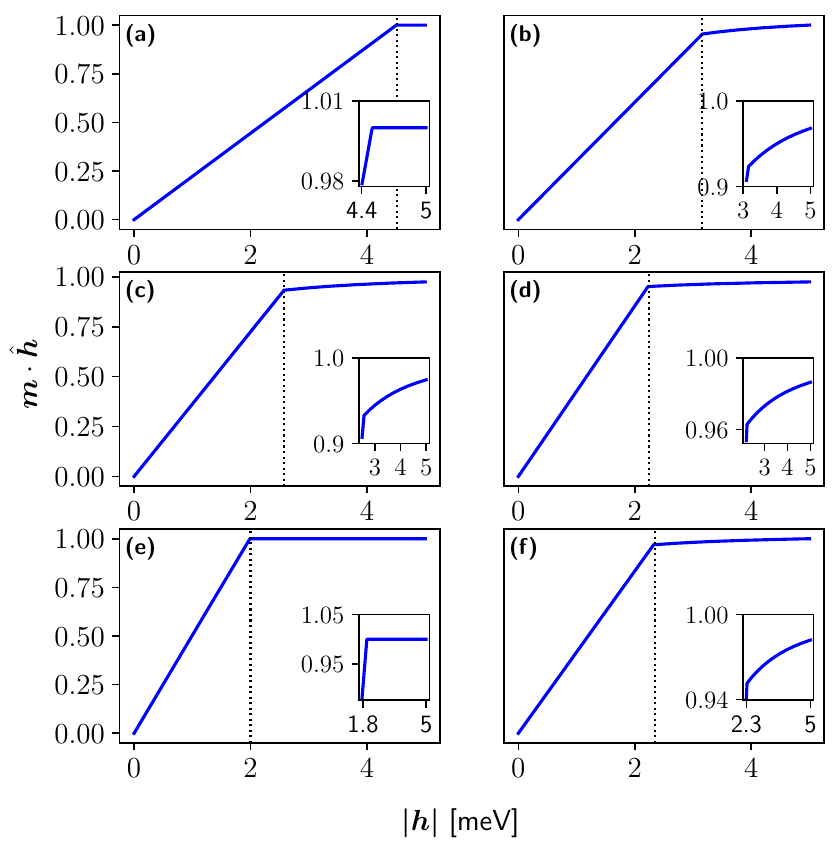}
    \caption{Classical magnetization curves calculated for various directions of the Zeeman field. (a) $(\theta_{{B}}, \phi_{B}) = (0, 0)$ corresponding to the $(1, 1, 1)$ direction in spin space. (b) $(\theta_{{B}}, \phi_{B}) = (30^\circ, 0)$. (c) $(\theta_{{B}}, \phi_{B}) = (45^\circ, 0)$. (d) $(\theta_{{B}}, \phi_{B}) = (60^\circ, 0)$. (e) $(\theta_{{B}}, \phi_{B}) = (90^\circ, 0)$. (f) In addition, we also show $(\theta_{{B}}, \phi_{B}) = (\cos^{-1}(\frac{1}{\sqrt{3}}), 180^\circ)$ which corresponds to $(0, 0, 1)$ direction in spin space. The insets for each representative field direction outline whether the system would be in a fully-polarized state (constant) or a high-field phase (asymptotic) with non-zero transverse magnetization.} 
    \label{fig: classical_magnetization}
\end{figure}

Fig.\,\ref{fig: classical_magnetization} outlines the classical magnetization curves evaluated for different representative field directions. The spherical angles $(\theta_{B}, \phi_{B})$ correspond to the zenith and azimuthal angles with respect to the honeycomb planes, respectively. Other than field directions of $\theta_{{B}} = 0$ (out-of-plane) and $\theta_{{B}} = 90^\circ$ (in-plane), the high-field phase does not exhibit complete polarization, i.e., finite transverse magnetization exists. That is, the classical spins are ferromagnetically aligned, yet they are not parallel to the external Zeeman field. This behavior has been observed for some of the ordered phases of the Kitaev-Heisenberg magnet\,\cite{spin_flop_1}.  

\begin{figure*}[t!]
    \centering
    \includegraphics[width=1 \linewidth]{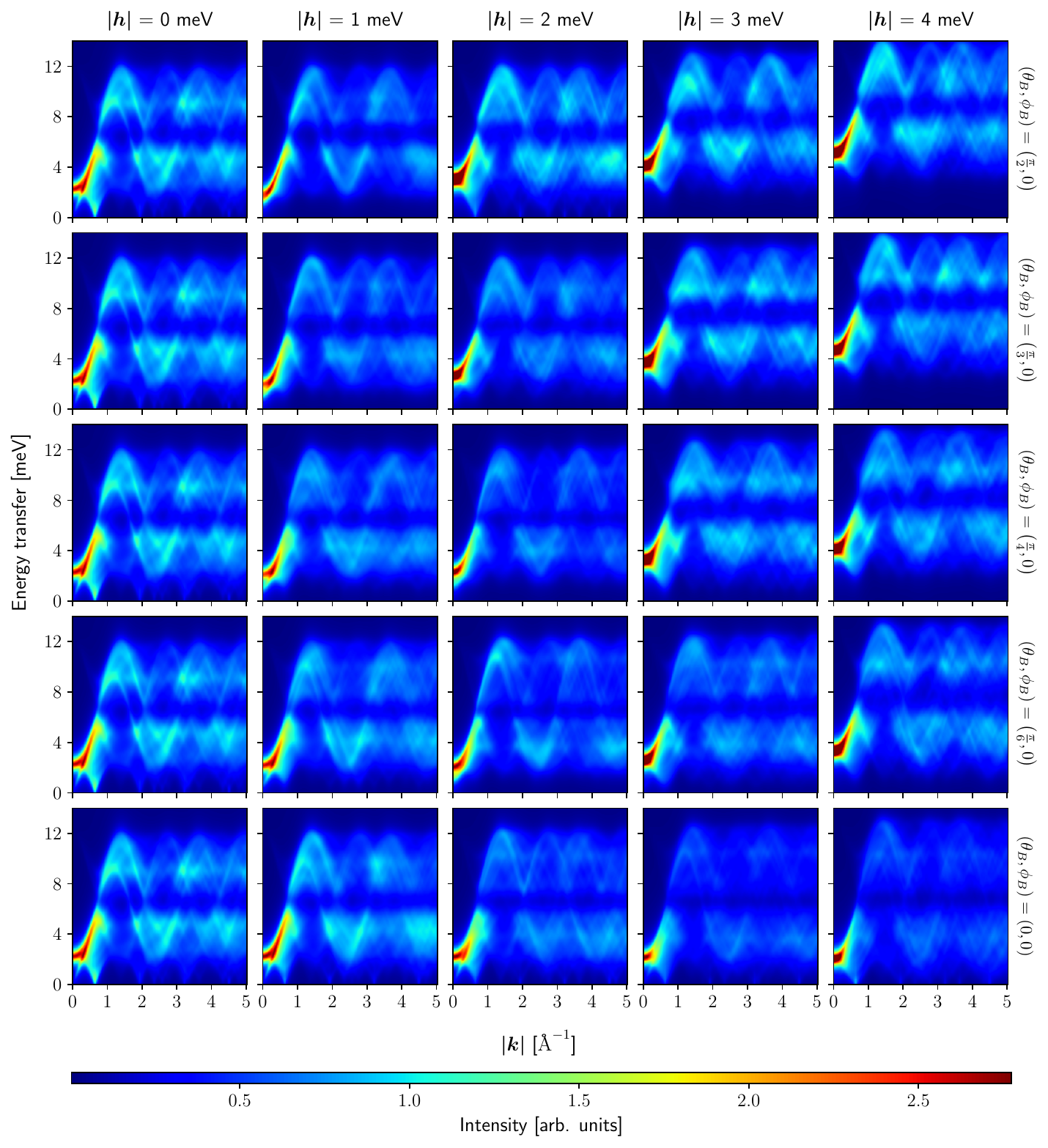}
    \caption{Powder-averaged DSFs computed for a range of applied magnetic field strengths. Each row corresponds to a fixed field orientation, specified by the angular pair $(\theta_{B},\phi_{B})$, while the columns illustrate the evolution of the spectra with increasing field magnitude.} 
    \label{fig: all_Bfield_DSF}
\end{figure*}

To assess the stability of the proposed magnetic unit cell, we employ a finite-size minimization procedure based on a gradual descent algorithm. This approach, however, does not guarantee convergence to the true ground state, as frustrated spin systems generically may exhibit a manifold of local minima at finite system sizes, even under periodic boundary conditions. One way of approaching this would be to randomize initial spin configurations and proceed with the iterative minimization for each of them. The corresponding minima would then be compared with the variationally minimized configuration. Based on our analysis, the variationally minimized ground state was stable compared to other local minima explored via this method.

\begin{figure}[t]
    \centering
    \includegraphics[width=1 \linewidth]{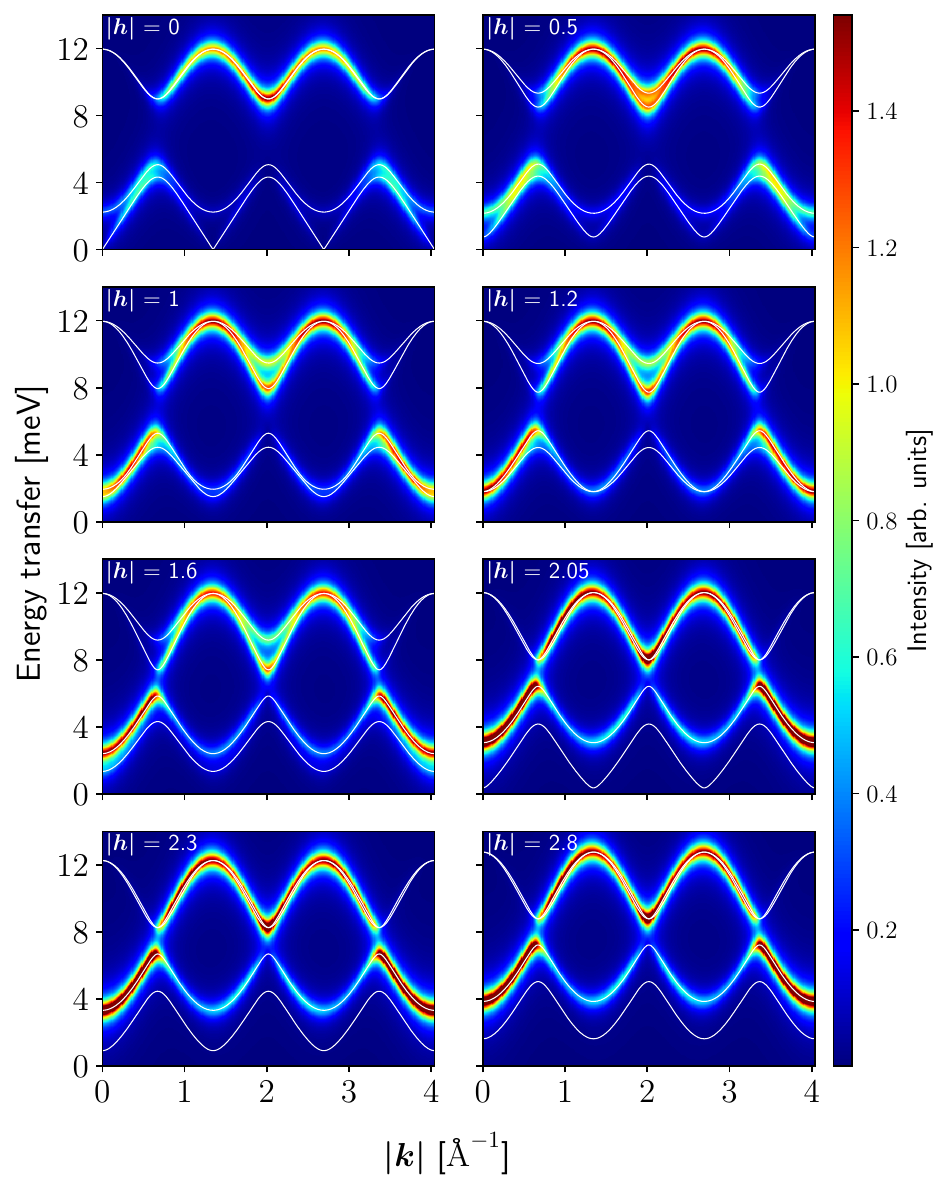}
    \caption{Field-dependent, momentum-resolved DSFs for various field strenghts $|\boldsymbol{h}|$. The shown $k$-path is parallel to 
    $\left(\sqrt{3}/2, 1/2, 0\right)$, i.e., as a purely 2D consideration it would correspond to $\Gamma \to M$.
    The direction of the Zeeman field is taken to be in-plane, $(\theta_{B}, \phi_{B}) = (\frac{\pi}{2}, 0)$ with its magnitude given in units of meV.} 
    \label{fig: band_touching_dsf}
\end{figure}

\begin{figure}[t]
    \centering
    \includegraphics[width=0.81 \linewidth]{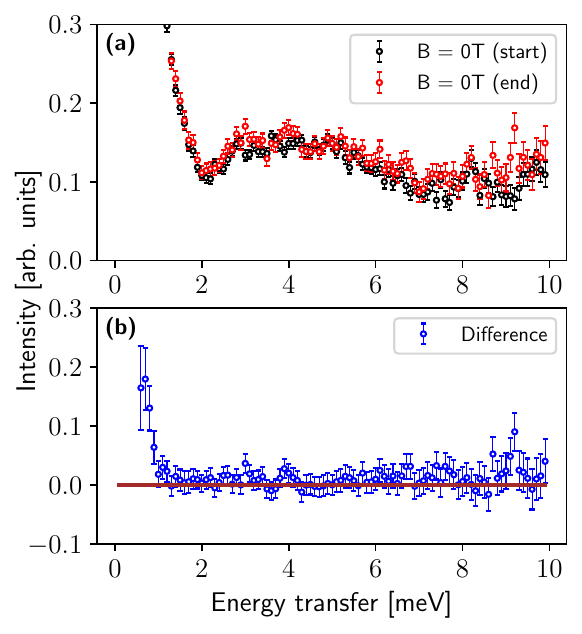}
    \caption{(a) Momentum-integrated INS spectra measured at zero applied magnetic field and after completion of the field-tuning cycle, illustrating whether the sample reorients itself after the application of an external field. (b) Corresponding difference map, obtained by subtracting the initial data from the final dataset.} 
    \label{fig: start_vs_end}
\end{figure}

\section{DSF under applied external fields}
\label{sec:appC}
As discussed in Sec.\,\ref{sec:in-field}, in order to investigate the evolution of the powder-averaged DSFs under different field orientations, we consider a range of candidate field directions and strengths. The corresponding classical ground-state configurations for these directions were determined in Appendix\,\ref{sec:appB}, which then serve as input for LSWT calculations of the DSFs. The resulting spectra, for both varying field directions and field strengths, are presented in Fig.\,\ref{fig: all_Bfield_DSF}.

For a magnetic field applied perpendicular to the honeycomb plane, i.e., $(\theta_{B},\phi_{B}) = \left(0, 0\right)$, the soft modes persist across the considered range of field strengths, and no gap opening is observed. In contrast, as the field direction deviates from the perpendicular orientation, i.e., $\theta_B\not= 0$, the soft modes progressively gap out, which correspond to the high-field phases of the classical magnetization curves shown in Fig.\,\ref{fig: classical_magnetization}. Another notable feature is the non-monotonic evolution of the soft modes with increasing field strength. Specifically, the soft modes transiently vanish, followed by a gap closing, and subsequently evolve into a robust gap opening at higher fields. This behavior is clearly illustrated in Fig.\,\ref{fig: band_touching_dsf}, where the magnon bands of a momentum-resolved DSF touch at a Zeeman field strength of approximately 1.2 meV, while a gap closes and re-opens at around 2.05 meV.

\section{Effects of magnetic field on sample reorientation}
\label{sec:appD}

A potential concern in powdered samples subjected to external magnetic fields is the possibility of field-induced sample reorientation. To assess this effect, we perform a full field sweep across the relevant range of field strengths, followed by a return to the initial conditions, i.e., $B=0$T. Fig.\,\ref{fig: start_vs_end} presents the momentum-integrated INS spectra measured at the beginning and at the end of the sweep cycle, along with their difference. The changes observed between the two datasets indicate that no significant reorientation occurs, and due to the smallness of the effect we can neglect it. 

\newpage

\bibliography{cobaltates}

\end{document}